\documentclass[11pt,a4paper,twoside]{article}
\usepackage[utf8]{inputenc}
\usepackage[english]{babel}
\usepackage{amsmath,amsfonts,amssymb,amsthm}
\usepackage{graphicx,hyperref}
\usepackage[margin=1in]{geometry} 
\usepackage{multirow}
\usepackage[justification=centering]{caption}
\usepackage{authblk,subcaption}
\usepackage{resizegather}

\theoremstyle{definition}
\newtheorem{defn}{Definition}[section]

\title{\textbf{The wrapped xgamma distribution for modeling circular data appearing in geological context}}

\author[a]{Hazem Al-Mofleh}
\author[b]{Subhradev Sen\thanks{E-mail:subhradev.stat@gmail.com}}

\affil[a]{\small{Department of Mathematics, Tafila Technical University, Tafila, Jordan.}}
\affil[b]{\small{Alliance School of Business, Alliance University, Bengaluru, India.}}

\begin{document}
\date{}
\maketitle
\begin{abstract}
The technique of \textit{wrapping} of a univariate probability distribution is very effective in getting a circular form of the underlying density. In this article, we introduce the circular (wrapped) version of xgamma distribution and study its different distributional properties. To estimate the unknown parameter, maximum likelihood method is proposed. A Monte-Carlo simulation study is performed to understand the behaviour of the estimates for varying sample size. To illustrate the application of the proposed distribution, a real data set on the long axis orientation of feldspar laths in basalt rock is analyzed and compared with other circular distributions.
\end{abstract}

\section{Introduction}
Measurements in direction is common in science and real life data observations. The popular encounter of such data sets might be related to direction of flight of a bird, orientation of certain animals, direction of magnetic field in a place, etc., could be two dimensional or three dimensional depending on the nature of measurements. Two dimensional directions are usually represented by angles measured with respect to some zero direction, in a view of rotational sense, clockwise or anti-clockwise. Such two-dimensional directional measurements with a circular representation, points on the circumference of a unit circle centered at the origin, is what is termed as \textit{circular data} (Rao and Sengupta, 2001). In a parametric sense, modeling of circular data sets is done based on the proper fit of a probability distribution, commonly known as a circular distribution.\\
Therefore, a circular distribution is a probability distribution of a random angle whose total probability is concentrated on the circumference of a unit circle. Any univariate probability distribution can be transformed to a circular distribution by a technique of ``wrapping'' and the resultant distribution, so obtained, is term as \textit{wrapped probability distribution}, for more details on wrapped distributions one could refer to L\'{e}vy (1939), Rao and Kozubowski (2004), Roy and Adnan (2012a, 2012b), Adnan and Roy (2014), Joshi and Jose (2018), and references therein.
 
The xgamma distribution is introduced and studied by Sen et al.(2016) in modeling lifetime data sets. Subsequent works on xgamma distribution, along with its variations and extensions, made the model a potential life distribution (see for more details, Sen and Chandra, 2017; Sen et al., 2017; Sen et al., 2018a; Sen et al., 2018b; Sen et al., 2018; Yadav et al., 2018; Maiti et al., 2018; Altun and Hamedani, 2018). In view of gaining popularity and importance of xgamma distribution, our aim in this article is to introduce the wrapped version of xgamma distribution, study the different properties, address the estimation process of distribution parameter and find suitable application in modeling circular data.

The rest of the article is organized as follows.
Section~\ref{sec2} introduces the wrapped version of xgamma distribution. Different distributional properties, such as, distribution function, characteristic function, trigonometric moments are studied in section~\ref{sec3} and in its dedicated subsections. Section~\ref{sec4} deals with estimation of parameter and a Monte-Carlo simulation study. Application in the context of data coming from geological field is illustrated in section~\ref{sec5}. Finally, section~\ref{sec6} concludes with major findings.
\section{Synthesis of wrapped xgamma distribution}
\label{sec2}
In this section we describe a method of synthesis for the circular version of xgamma distribution following the technique of wrapping of univarite density.\\
If $X$ follows xgamma distribution (Sen et al, 2016) with parameter $\lambda (>0)$, then the probability density function (pdf) of xgamma distribution is given by
\begin{align}
f(x)=\frac{\lambda^2}{(1+\lambda)}\left(1+\frac{\lambda}{2}x^2\right)e^{-\lambda x}\quad \text{for}\quad x>0.
\end{align}
We denoted it by $X\sim XG(\lambda)$.\\
The corresponding cumulative density function (cdf) is given by
\begin{align}
\label{cdfxg}
F(x)=1-\frac{\left(1+\lambda+\lambda x+\frac{\lambda^2}{2}x^2\right)}{(1+\lambda)}e^{-\lambda x}\quad \text{for} \quad x>0.
\end{align}
Now, the circular (wrapped) xgamma random variable is defined as $\theta \equiv X (\;\text{mod}\;2\pi)$, such that for $\theta \in [0, 2\pi)$, the pdf can be derived as
\begin{align*}
g(\theta)&=\sum_{k=0}^{\infty}f(\theta+2k\pi)
=\frac{\lambda^2}{(1+\lambda)}\sum_{k=0}^{\infty}e^{-\lambda(\theta+2k \pi)}\left[1+\frac{\lambda}{2}(\theta+2k\pi)^2\right]\\
&=\frac{\lambda^2 e^{-\lambda \theta}}{(1+\lambda)}\left[\sum_{k=0}^{\infty}e^{-2\pi \lambda k}+\frac{\lambda}{2}\sum_{k=0}^{\infty}e^{-2\pi \lambda k}(\theta+2k\pi)^2\right]\\
&=\frac{\lambda^2 e^{-\lambda \theta}}{(1+\lambda)}\left[\left(1+\frac{\lambda \theta^2}{2}\right)\sum_{k=0}^{\infty}e^{-2\pi \lambda k}+2\pi \theta \lambda\sum_{k=0}^{\infty}k e^{-2\pi \lambda k}+2 \pi^2 \lambda \sum_{k=0}^{\infty}k^{2} e^{-2\pi \lambda k} \right]\\
&=\frac{\lambda^2 e^{-\lambda \theta}}{(1+\lambda)}\left[\left(1+\frac{\lambda\theta ^2}{2}\right)\frac{1}{(1-e^{-2\pi \lambda})}+\frac{2\pi \theta \lambda e^{-2\pi \lambda}}{(1-e^{-2 \pi \lambda})^2}+\frac{2 \pi^2 \lambda e^{2 \pi \lambda} \left( e^{2 \pi \lambda}+1 \right)}{\left(e^{2 \pi \lambda}-1\right)^3}\right]\\
&=\frac{\lambda ^2 e^{-\theta  \lambda }}{(\lambda +1) \left(1-e^{-2 \pi  \lambda }\right)}\left[\left(1+\frac{\lambda\theta ^2}{2}\right)+2 \pi \lambda  \left((\pi -\theta )e^{-2 \pi  \lambda } +(\theta +\pi )\right)\frac{e^{-2 \pi  \lambda }}{\left(1-e^{-2 \pi  \lambda }\right)^2}\right].
\end{align*}

\begin{figure}[h!]
\begin{subfigure}{.5\textwidth}
  \centering
  \includegraphics[width=.9\linewidth]{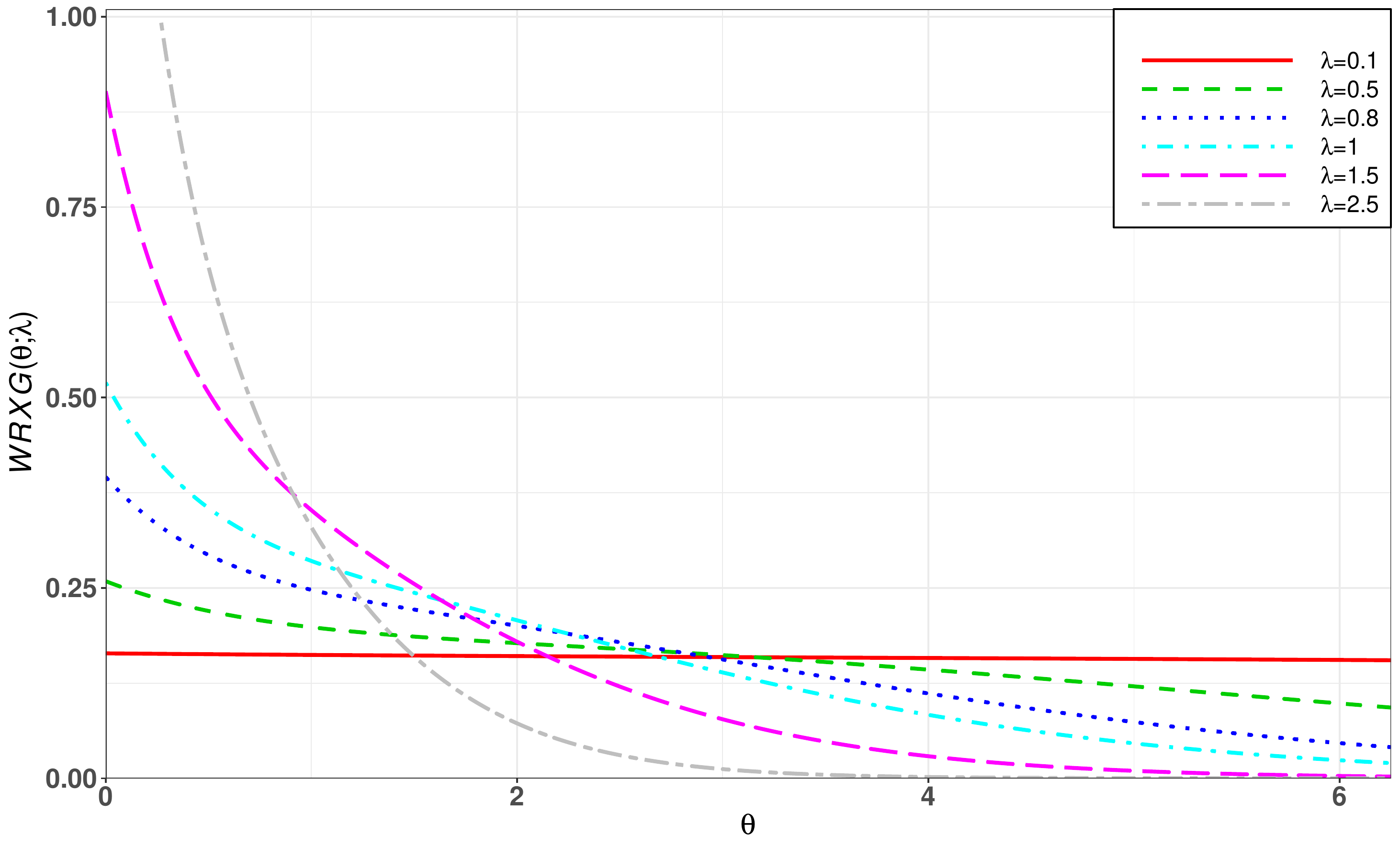}
  \caption{Probability density function}
  \label{fig:pdf1}
\end{subfigure}%
\begin{subfigure}{.5\textwidth}
  \centering
  \includegraphics[width=.9\linewidth]{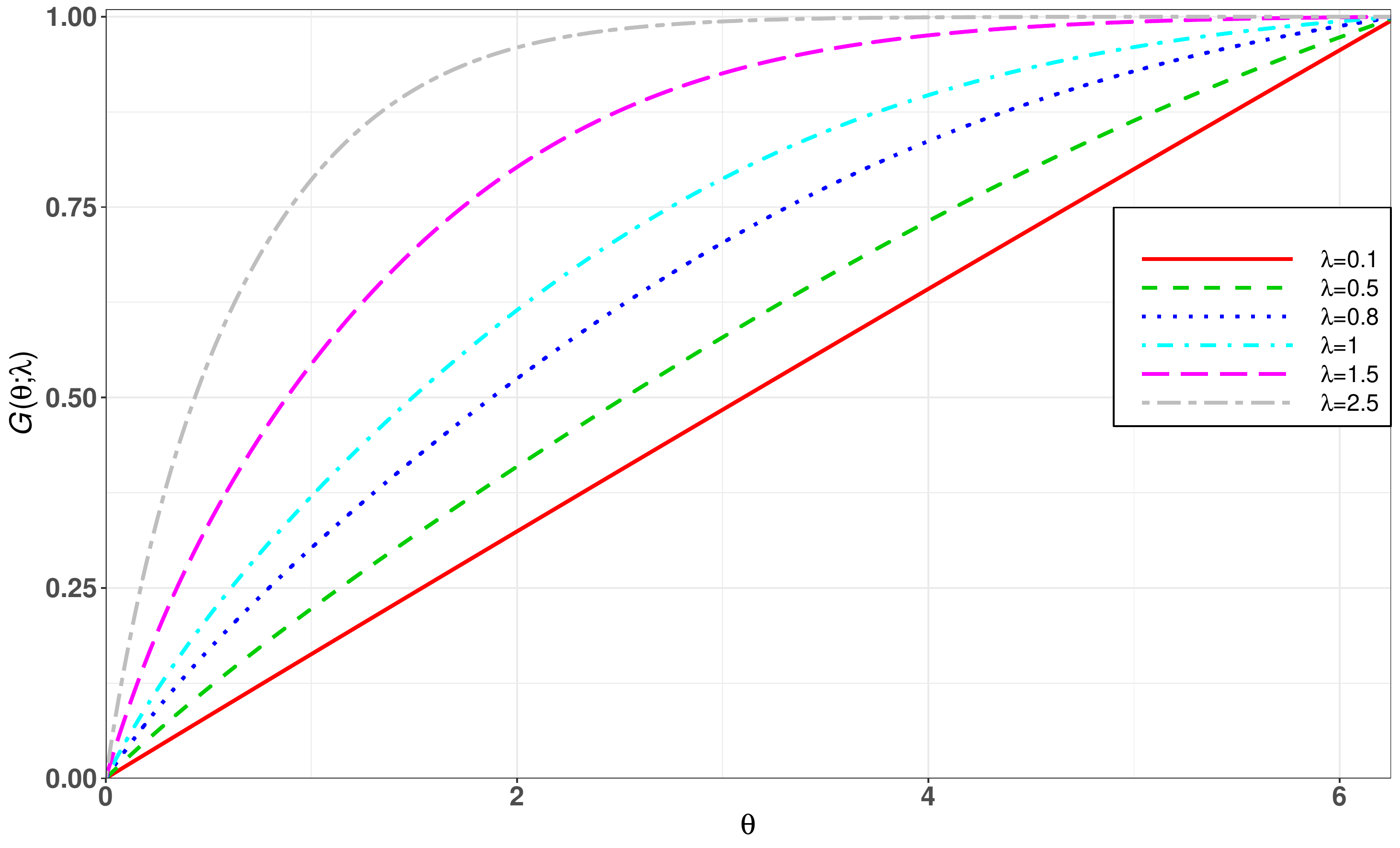}
  \caption{Cumulative distribution function}
  \label{fig:cdf1}
\end{subfigure}
\caption{Plots of $WRXG(\theta;\lambda)$ for selected values of parameter.}
\label{fig:denplot}
\end{figure}

\begin{figure}[h!]
\begin{subfigure}{.5\textwidth}
  \centering
  \includegraphics[width=.9\linewidth]{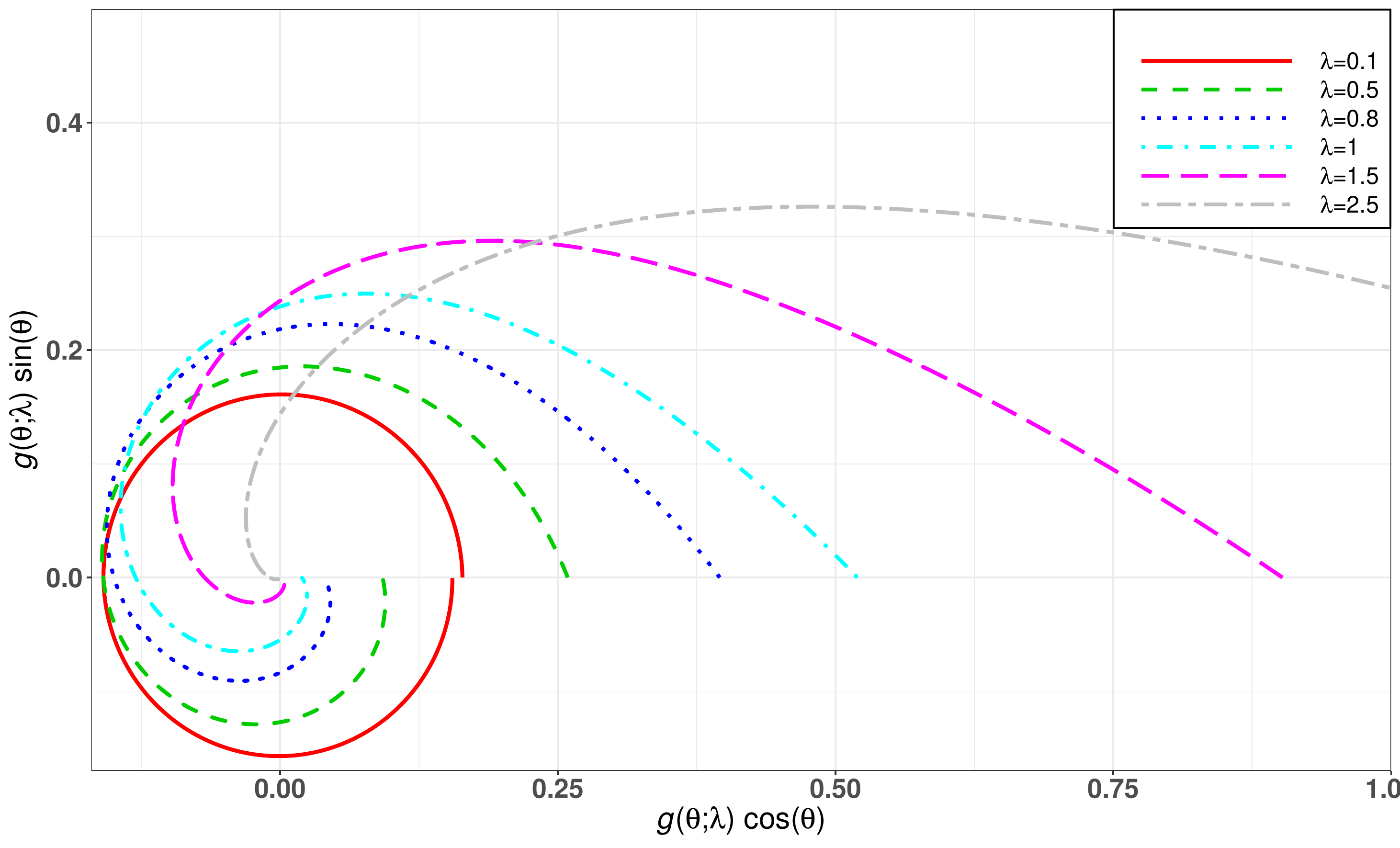}
  \caption{Probability density function}
  \label{fig:pdf2}
\end{subfigure}%
\begin{subfigure}{.5\textwidth}
  \centering
  \includegraphics[width=.9\linewidth]{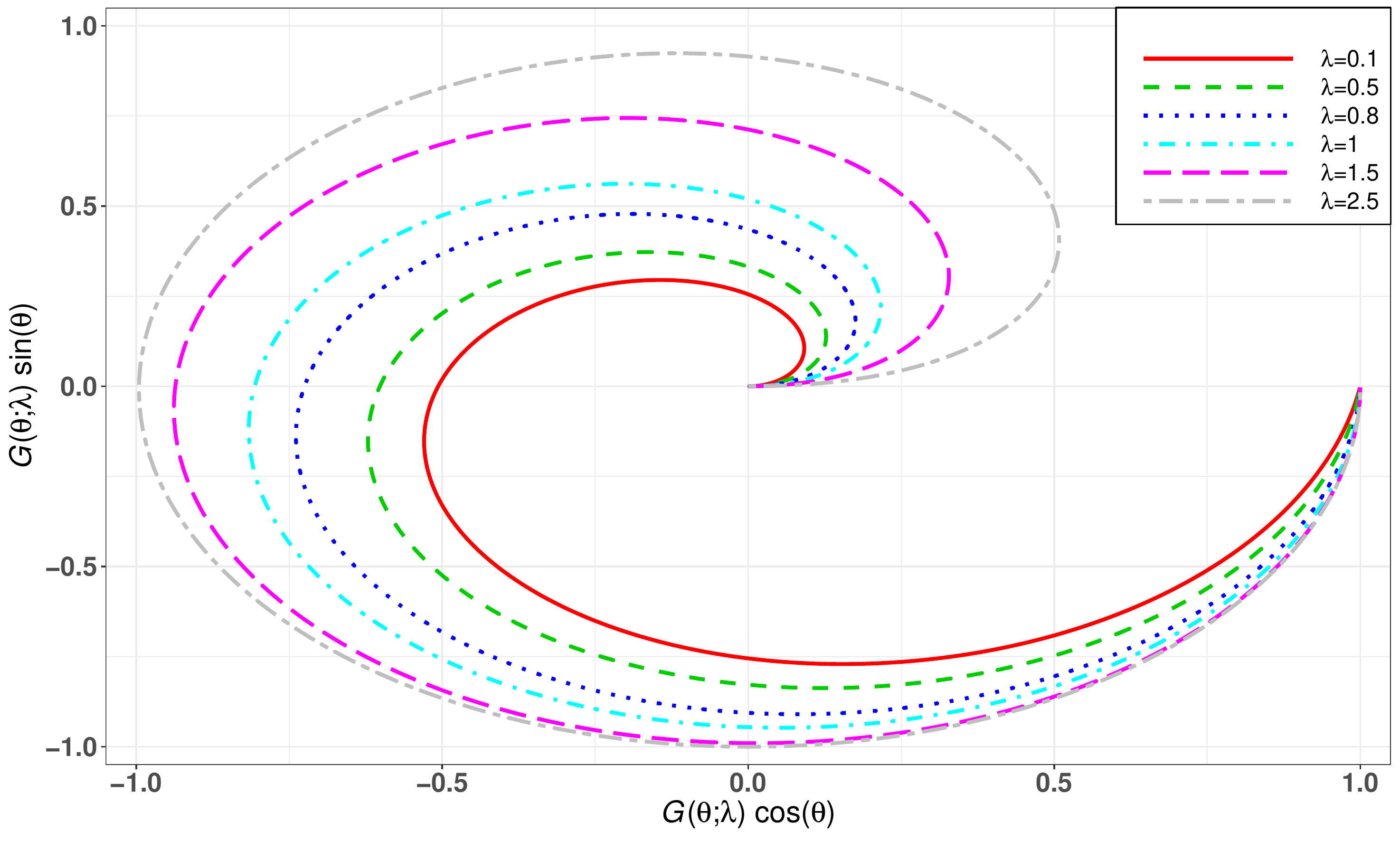}
  \caption{Cumulative distribution function}
  \label{fig:cdf2}
\end{subfigure}
\caption{Circular representation of $WRXG(\theta;\lambda)$ for different values of parameter.}
\label{fig:circplot1}
\end{figure}

\begin{figure}[h!]
\begin{subfigure}{.5\textwidth}
  \centering
  \includegraphics[width=.9\linewidth]{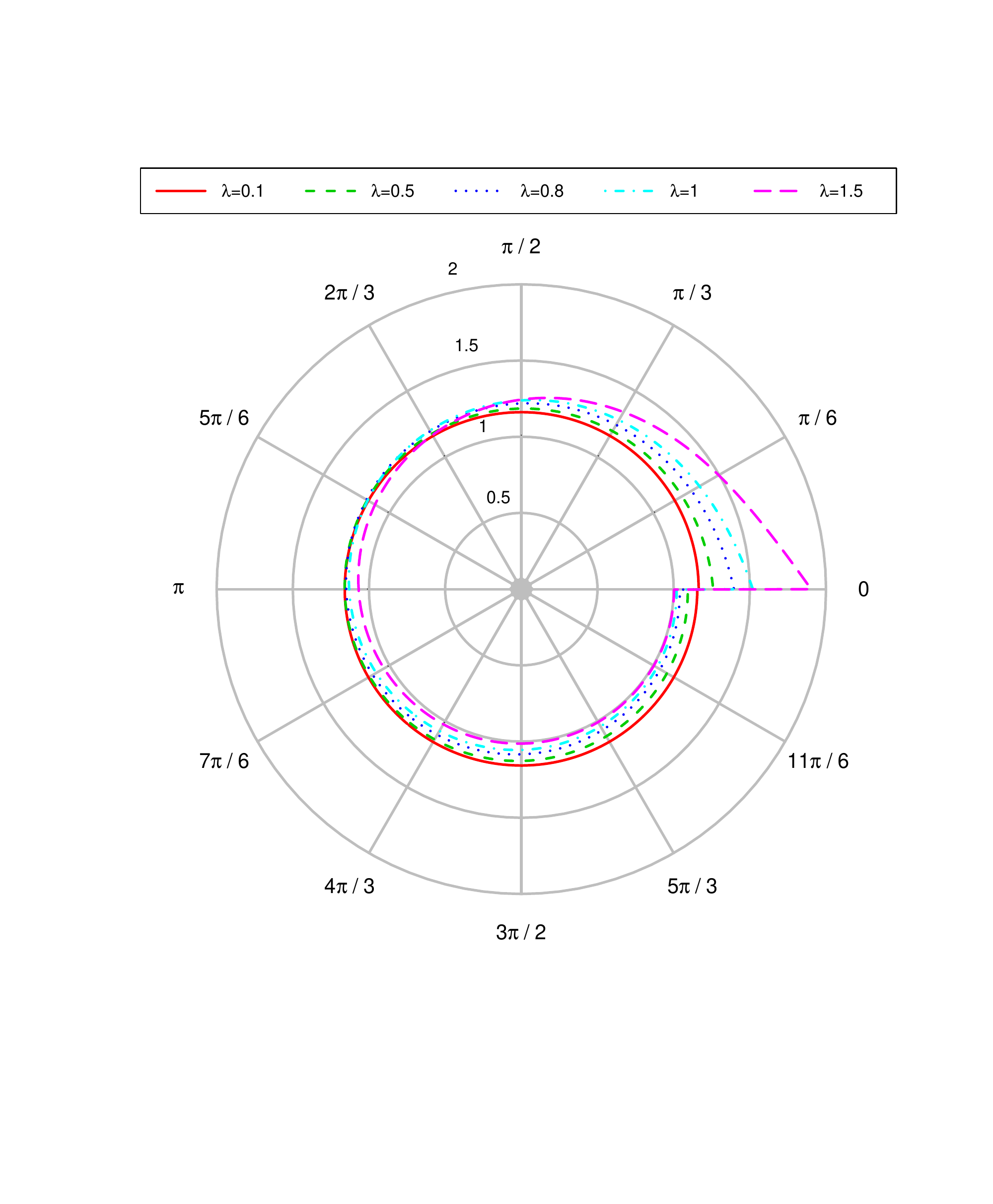}
  \caption{Probability density function}
  \label{fig:pdf3}
\end{subfigure}%
\begin{subfigure}{.5\textwidth}
  \centering
  \includegraphics[width=.9\linewidth]{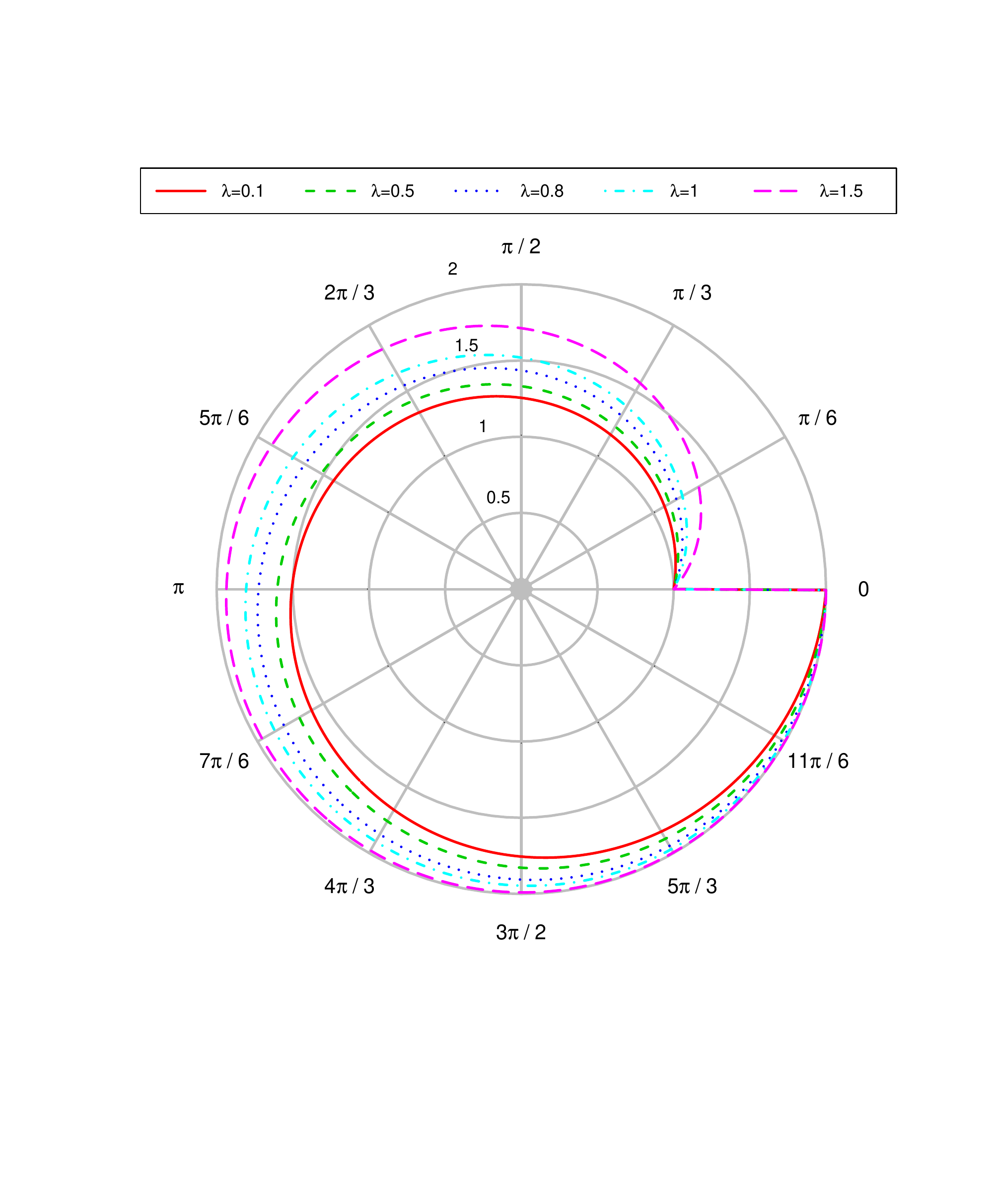}
  \caption{Cumulative distribution function}
  \label{fig:cdf3}
\end{subfigure}
\caption{Circular $WRXG(\theta;\lambda)$ for different values of parameter.}
\label{fig:circplot2}
\end{figure}

We have the following definition for wrapped xgamma distribution.
\begin{defn}
A circular random variable $\theta$ is said to follow the wrapped xgamma (WRXG) distribution  with parameter $(\lambda>0)$ if its pdf is of the form
\begin{align}\label{pdf1}
g(\theta)=\frac{\lambda ^2 e^{-\theta  \lambda }}{(\lambda +1) \left(1-e^{-2 \pi  \lambda }\right)}\left[1+\frac{\lambda\theta ^2}{2}+2 \pi \lambda  \left((\pi -\theta )e^{-2 \pi  \lambda } +(\theta +\pi )\right)\frac{e^{-2 \pi  \lambda }}{\left(1-e^{-2 \pi  \lambda }\right)^2}\right]; \theta \in [0,2\pi).
\end{align}
It is denoted by $\theta \sim WRXG(\lambda)$.
\end{defn}
Figures \ref{fig:denplot} to \ref{fig:circplot2} show the plots of wrapped xgamma distribution for different values of $\lambda$.

\section{Distributional properties}
\label{sec3}
In this section, we study different properties, such as, cumulative distribution function, characteristic function, trigonometric moments and other related measures of $WRXG(\lambda)$.

\subsection{Cumulative distribution function}
Following Mardia and Jupp (2000) and Rao and Sengupta (2001), the cdf of $WRXG(\theta;\lambda)$ can be obtained as follows.
\begin{align}
\notag G(\theta)&=\sum_{k=0}^{\infty}\left[F(\theta+2 \pi k)-F(2 \pi k)\right] \\ \notag
&=\sum_{k=0}^{\infty} \frac{e^{-2 \pi k \lambda}}{(1+\lambda)} \left[
{1+\lambda+2 \pi k \lambda+\frac{\lambda^2}{2}\left(2 \pi k\right)^2}
-{\left(1+\lambda+\lambda \left(\theta+2 \pi k\right)+\frac{\lambda^2}{2}\left(\theta+2 \pi k\right)^2\right)}e^{-\theta \lambda}\right]\\ \notag
&=\left(1-\frac{1+\lambda(1+ \theta)+ \frac{1}{2} \theta ^2 \lambda ^2}{\lambda +1}e^{-\theta  \lambda }\right)\sum_{k=0}^{\infty} e^{-2\pi k \lambda}+\frac{2 \pi \lambda }{\lambda +1}\left(1-(1+\theta  \lambda ) e^{-\theta  \lambda }\right)\sum_{k=0}^{\infty} k e^{-2\pi k \lambda}\\  \notag
& \ \ \ \ +\frac{2 \pi ^2 \lambda ^2}{\lambda +1} \left( 1-e^{-\theta  \lambda } \right)\sum_{k=0}^{\infty} {k^2} e^{-2\pi k \lambda}\\ \notag
&=\left(1-\frac{1+\lambda(1+ \theta)+ \frac{1}{2} \theta ^2 \lambda ^2 }{\lambda +1}e^{-\theta  \lambda }\right)\frac{1}{1-e^{-2 \pi  \lambda }}
+\frac{2 \pi \lambda }{\lambda +1}\left(1-(1+\theta  \lambda) e^{-\theta  \lambda }\right)\frac{e^{-2 \pi  \lambda }}{\left(1-e^{-2 \pi  \lambda }\right)^2}\\
& \ \ \ \ +\frac{2 \pi ^2 \lambda ^2}{\lambda +1} \left( 1-e^{-\theta  \lambda } \right)\frac{e^{-2 \pi  \lambda } \left(e^{-2 \pi  \lambda }+1\right)}{\left(1-e^{-2 \pi  \lambda }\right)^3}. 
\end{align}

\subsection{Characteristic function}
By Rao and Sengupta (2001), the trigonometric moments of order $p$ for a wrapped
circular distribution corresponds to the value of the characteristic function of the unwrapped random variable $X$, say $\phi_{X}(t)$ at the integer value $p$, i.e., $\varphi_{\theta}(p)=\phi_{X}(p)$. The characteristic function of xgamma distribution is given by (Sen et al., ~2018a)
\begin{align*}
\phi_{X}(t)=E\left(e^{i t X}\right) 
=\frac{\lambda ^2}{(\lambda +1)}(\lambda -i t)^{-3}\left[(\lambda ^2+\lambda -t^2){-i 2 t \lambda}\right],\; i=\sqrt{-1}.
\end{align*}
The characteristic function of wrapped xgamma distribution is given by
\begin{align*}
\varphi_{\theta}(p)=E\left(e^{i p \theta}\right) 
&=\frac{\lambda ^2}{(\lambda +1)}(\lambda -i p)^{-3}\left((\lambda ^2+\lambda -p^2){-i (2 p \lambda)}\right),
\end{align*}
where $p=\pm 1, \pm 2,....$
Since any complex number to the power $r$, $(a-ib)^r$ for $a,b, r\in R$ can be written using Euler's formula $\left(\sqrt{a^2+b^2}\right)^{r} \ e^{i r \tan^{-1}(b/a)}$, so $(\lambda -i p)^{-3}$ and $\left((\lambda ^2+\lambda -p^2){-i (2 p \lambda)}\right)$ can be written as
\begin{align*}
(\lambda -i p)^{-3}=\left(\sqrt{\lambda^2+p^2}\right)^{-3} \ e^{i 3 \tan^{-1}(p/\lambda)} \ \ \ \ \ \ \ \ \ \ \ \ \ \ \ \ \ \ \ \ \ \ \ \ \ \ \ \\
\left((\lambda ^2+\lambda -p^2){-i (2 p \lambda)}\right)=\sqrt{(\lambda ^2+\lambda -p^2)^2+4 p^2 \lambda^2} \ e^{-i \tan^{-1}\left(2 p \lambda/(\lambda ^2+\lambda -p^2)\right)}
\end{align*}
Therefore, $\varphi(p)$ becomes
\begin{align}\label{eq5}
\varphi_{\theta}(p)&= \frac{\lambda ^2}{(\lambda +1)}\left(\sqrt{\lambda^2+p^2}\right)^{-3} \sqrt{(\lambda ^2+\lambda -p^2)^2+4 p^2 \lambda^2} \ e^{i\left[ 3 \tan^{-1}(p/\lambda)- \tan^{-1}\left(2 p \lambda/(\lambda ^2+\lambda -p^2)\right)\right]}. \notag \\ 
&=\frac{\lambda ^2}{(\lambda +1)} \sqrt{\frac{(\lambda ^2+\lambda -p^2)^2+4 p^2 \lambda^2}{\left({\lambda^2+p^2}\right)^{3}} }\ e^{i\left[ 3 \tan^{-1}(p/\lambda)- \tan^{-1}\left(2 p \lambda/(\lambda ^2+\lambda -p^2)\right)\right]}.
\end{align}

\subsection{Trigonometric moments and related measures}
By Rao and Sengupta (2001), $\varphi_{\theta}(p)=\rho_{p} e^{i\mu_{p}}$. Now, from (\ref{eq5}) we have
\begin{align}\label{eq6}
\rho_{p}=\frac{\lambda ^2}{(\lambda +1)} \sqrt{\frac{(\lambda ^2+\lambda -p^2)^2+4 p^2 \lambda^2}{\left({\lambda^2+p^2}\right)^{3}} }.
\end{align}
and
\begin{align}\label{eq7}
\mu_{p}=3 \tan^{-1}(p/\lambda)- \tan^{-1}\left(2 p \lambda/(\lambda ^2+\lambda -p^2)\right).
\end{align}
The mean direction, the resultant length, the circular variance and the circular standard deviation, respectively, are given by
\begin{align}\label{eq8}
\mu=\mu_{1}=3 \tan^{-1}(1/\lambda)- \tan^{-1}\left(2\lambda/(\lambda ^2+\lambda -1)\right),
\end{align}
\begin{align}\label{eq9}
\rho=\rho_{1}=\frac{\lambda ^2}{(\lambda +1)} \sqrt{\frac{(\lambda ^2+\lambda -1)^2+4 \lambda^2}{\left({\lambda^2+1}\right)^{3}} },
\end{align}
\begin{align}\label{eq10}
V_{0}=1-\rho_{1}=1-\frac{\lambda ^2}{(\lambda +1)} \sqrt{\frac{(\lambda ^2+\lambda -1)^2+4 \lambda^2}{\left({\lambda^2+1}\right)^{3}} }
\end{align}
and
\begin{align}\label{eq11}
\sigma_{0}=\sqrt{-2\log(1-V_{0})}=\sqrt{-2\log\left(\frac{\lambda ^2}{(\lambda +1)} \sqrt{\frac{(\lambda ^2+\lambda -1)^2+4 \lambda^2}{\left({\lambda^2+1}\right)^{3}} }\right)}
\end{align}
By the definition of trigonometric moments and since any complex number can be written as $\alpha_{p}+i\beta_{p}$, $\varphi_{\theta}(p)$ in (\ref{eq5}) can be rewritten as 
$\varphi_{\theta}(p)=\alpha_{p}+i\beta_{p}$ where
\begin{align}\label{eq12}
\alpha_{p}&=\rho_{p}\cos(\mu_{p})  \notag \\
&=\frac{\lambda ^2}{(\lambda +1)} \sqrt{\frac{(\lambda ^2+\lambda -p^2)^2+4 p^2 \lambda^2}{\left({\lambda^2+p^2}\right)^{3}}} \cos(3 \tan^{-1}(p/\lambda)- \tan^{-1}\left(2 p \lambda/(\lambda ^2+\lambda -p^2)\right)),
\end{align}
and
\begin{align}\label{eq13}
\beta_{p}&=\rho_{p}\sin(\mu_{p})  \notag \\
&=\frac{\lambda ^2}{(\lambda +1)} \sqrt{\frac{(\lambda ^2+\lambda -p^2)^2+4 p^2 \lambda^2}{\left({\lambda^2+p^2}\right)^{3}}} \sin(3 \tan^{-1}(p/\lambda)- \tan^{-1}\left(2 p \lambda/(\lambda ^2+\lambda -p^2)\right)),
\end{align}
where $p=0, \pm 1, \pm 2,...$. \\
$\alpha_{p}$ and $\beta_{p}$ are called the non central trigonometric moments of the distribution.
The central trigonometric moments are defined as $\bar{\alpha_{p}}=\rho_{p}\cos(\mu_{p}-p\mu_{1})$ and $\bar{\beta_{p}}=\rho_{p}\sin(\mu_{p}-p\mu_{1})$. \\
Hence, the central trigonometric moments of the respective distribution are given
by
\begin{align}\label{eq14}
\bar{\alpha_{p}}=\frac{\lambda ^2}{(\lambda +1)} \sqrt{\frac{(\lambda ^2+\lambda -p^2)^2+4 p^2 \lambda^2}{\left({\lambda^2+p^2}\right)^{3}}} \cos\left(\kappa_{\lambda,p}\right).
\end{align}
and
\begin{align}\label{eq15}
\bar{\beta_{p}}=\frac{\lambda ^2}{(\lambda +1)} \sqrt{\frac{(\lambda ^2+\lambda -p^2)^2+4 p^2 \lambda^2}{\left({\lambda^2+p^2}\right)^{3}}} \sin\left(\kappa_{\lambda,p}\right),
\end{align}
where \\
$\kappa_{\lambda,p}=\tan^{-1}(p/\lambda)-3p\tan^{-1}(1/\lambda)+p\tan^{-1}\left(2\lambda/(\lambda ^2+\lambda -1)\right)- \tan^{-1}\left(2 p \lambda/(\lambda ^2+\lambda -p^2)\right)$. \\
The coefficient of skewness, $\zeta_1^0=\frac{\bar{\beta}_{2}}{V_0^{3/2}}$ and the coefficient of kurtosis, $\zeta_2^0=\frac{\bar{\alpha}_{2}-(1-V_{0})^4}{V_0^{2}}$, respectively, are given by
\begin{equation}
\zeta_1^0=\frac{\frac{\lambda^2}{(\lambda +1)} \sqrt{\frac{(\lambda ^2+\lambda-4)^2+16\lambda^2}{\left({\lambda^2+4}\right)^{3}}} \sin\left(\kappa_{\lambda,2}\right)}{\left(1-\frac{\lambda ^2}{(\lambda +1)} \sqrt{\frac{(\lambda ^2+\lambda -1)^2+4 \lambda^2}{\left({\lambda^2+1}\right)^{3}} }\right)^{3/2}}
\end{equation}
and
\begin{equation}
\zeta_2^0=\frac{\frac{\lambda^2}{(\lambda +1)} \sqrt{\frac{(\lambda ^2+\lambda-4)^2+16\lambda^2}{\left({\lambda^2+4}\right)^{3}}} \cos\left(\kappa_{\lambda,2}\right)-\left(\frac{\lambda ^2}{(\lambda +1)} \sqrt{\frac{(\lambda ^2+\lambda -1)^2+4 \lambda^2}{\left({\lambda^2+1}\right)^{3}} }\right)^{4}}{\left(1-\frac{\lambda ^2}{(\lambda +1)} \sqrt{\frac{(\lambda ^2+\lambda -1)^2+4 \lambda^2}{\left({\lambda^2+1}\right)^{3}} }\right)^{2}},
\end{equation}
where \\
$\kappa_{\lambda,2}=\tan^{-1}(2/\lambda)-6\tan^{-1}(1/\lambda)+2\tan^{-1}\left(2\lambda/(\lambda ^2+\lambda -1)\right)- \tan^{-1}\left(4\lambda/(\lambda ^2+\lambda -4)\right)$. \\

\begin{table}[!h]
	\centering
	\caption{Values of different characteristics of wrapped xgamma distribution for varying $\lambda$.}
	\label{tab1}
	\resizebox{\textwidth}{!}{
\begin{tabular}{llcccccc}
\hline
\multicolumn{2}{l}{\multirow{2}{*}{Characteristics of WRXG distribution}} & \multicolumn{6}{c}{$\lambda$}                                   \\ 
\multicolumn{2}{c}{}                                                                & 0.1      & 0.7      & 1        & 2.5      & 4        & 8        \\ \hline
Mean direction                                   & $\mu$                            & 4.63443  & 1.44430  & 1.24905  & 0.56855  & 0.33641  & 0.15142  \\
Resultant length                                 & $\rho$                           & 0.00817  & 0.22390  & 0.39528  & 0.84367  & 0.94118  & 0.98760  \\
Circular variance                                & $V_{0}$                          & 0.99183  & 0.77610  & 0.60472  & 0.15633  & 0.05882  & 0.01240  \\
Circular standard deviation                      & $\sigma_{0}$                     & 3.10074  & 1.73005  & 1.36250  & 0.58308  & 0.34821  & 0.15795  \\ \hline
\multirow{4}{*}{Non-central trigonometric moments}                & $\alpha_{1}$                     & -0.00064 & 0.02825  & 0.12500  & 0.71095  & 0.88842  & 0.97630  \\
                                                 & $\alpha_{2}$                     & -0.00021 & -0.02698 & -0.05600 & 0.37595  & 0.66560  & 0.91187  \\
                                                 & $\beta_{1}$                      & -0.00815 & 0.22211  & 0.37500  & 0.45425  & 0.31069  & 0.14897  \\
                                                 & $\beta_{2}$                      & -0.00442 & -0.11711 & -0.19200 & 0.47073  & 0.46080  & 0.27718  \\ \hline
\multirow{4}{*}{Central trigonometric moments}                    & $\bar{\alpha}_{1}$               & 0.00817  & 0.22390  & 0.39528  & 0.84367  & 0.94118  & 0.98760  \\
                                                 & $\bar{\alpha}_{2}$               & 0.00443  & -0.11958 & -0.19986 & 0.57025  & 0.78040  & 0.94324  \\
                                                 & $\bar{\beta}_{1}$                & 0        & 0        & 0        & 0        & 0        & 0        \\
                                                 & $\bar{\beta}_{2}$                & 0.00014  & 0.01199  & -0.00759 & 0.19426  & 0.21525  & 0.13646  \\ \hline
Coefficient of skewness                          & $\zeta_1^0$                      & 0.00446  & 0.17571  & 0.39809  & -2.31898 & -3.81490 & -5.34250 \\
Coefficient of kurtosis                          & $\zeta_2^0$                      & -0.00049 & -0.00947 & -0.25928 & 3.21185  & 6.66040  & 11.11000 \\ \hline
\end{tabular}}
\end{table}
For different values of $\lambda$, table~\ref{tab1} shows the values of these characteristics of the wrapped xgamma distribution.

\section{Parameter estimation}
\label{sec4}
In this section, we consider the Maximum likelihood method to estimate the unknown parameter of
WRXG distribution for a complete sample situation. Let $\theta_{1},\theta_{2},\ldots ,\theta_{n}$ be a random
sample of size $n$ from $WRXG(\lambda)$. The log-likelihood function for $\lambda$ reduces to%

\begin{eqnarray*}
\ell(\lambda)  &=&2n\log(\lambda)-{n \theta  \lambda }-n\log{(\lambda +1)}-n\log{\left(1-e^{-2 \pi  \lambda }\right)}\\
&&+n\log{\left[\left(1+\frac{\lambda\theta ^2}{2}\right)+2 \pi \lambda  \left((\pi -\theta )e^{-2 \pi  \lambda } +(\theta +\pi )\right)\frac{e^{-2 \pi  \lambda }}{\left(1-e^{-2 \pi  \lambda }\right)^2}\right]}.
\end{eqnarray*}%
The Maximum likelihood estimators (MLEs) of the unknown parameter $\lambda$ of the WRXGD distribution can be obtained by
maximizing the last equation. This can also be done by using different
programs namely \texttt{R} (\texttt{optim} function), \texttt{SAS} (\texttt{PROC NLMIXED}) or by solving the nonlinear likelihood equation obtained by
differentiating $\ell$ with respect to $\lambda$ and equal it to zero, one of the methods can be used, the Newton-Rapshon as a numerical method to find the solution of the nonlinear equation.

\subsection{A simulation study}
To study the performance of the WRXG distribution and to investigate the behavior of maximum likelihood estimator, we generate 10,000 samples of the WRXG distribution, which sample sizes, $n = \{30,80,100,200,350\}$, and by choosing
$\widehat{\lambda}=(0.1,0.7,1,2.5,4,8)$, for each parameters combination and each sample, we evaluate: the average of absolute value of biases ($|Bias(\widehat{\lambda})|$), 
$$|Bias(\widehat{\lambda})|=\frac{1}{N}\sum_{i=1}^{N}|\widehat{\lambda}-\lambda|,$$  
the mean square error of the estimates ($MSE(\widehat{\lambda})$), 
$$MSE(\widehat{\lambda})=\frac{1}{N}\sum_{i=1}^{N}(\widehat{\lambda}-\lambda)^2,$$
and the mean relative estimates ($MRE(\widehat{\lambda})$), 
$$MRE(\widehat{\lambda})=\frac{1}{N}\sum_{i=1}^{N}|\widehat{\lambda}-\lambda|/\lambda.$$
\textsf{R} software (version 3.5.2) (R Core Team, 2018), is been utilized for calculations involved in the simulation study.\\
Table~\ref{tab:2} displays $|Bias(\widehat{\lambda})|$, $MSE(\widehat{\lambda})$, and $MRE(\widehat{\lambda})$ of the MLEs, it can be observed that the values of $|Bias(\widehat{\lambda})|$, $MSE(\widehat{\lambda})$ and $MRE(\widehat{\lambda})$ decrease as sample size increases. From the results in Table~\ref{tab:2} we can conclude that the Estimation method, viz. MLE preserves the consistency property for all combinations of parameter.
\begin{table}[!h]
\begin{center}
\caption{Result of simulation study showing estimates, bias, MSE, and MRE \\for varying $\lambda$ and $n$.}
\label{tab:2}
\begin{tabular}{lccccccc}
\hline
 & &\multicolumn{6}{c}{$\lambda$}\\
$n\downarrow$ &Measure &0.1 &0.7 &1.0 &2.5 &4.0 &8.0\\
\hline
\multirow{4}{*}{30} &$\widehat{\lambda}$ &0.18013 &0.67309 &1.01357 &2.55849 &4.11511 &8.28518 \\
 &$|Bias(\widehat{\lambda})|$ &0.17816 &0.17014 &0.13340 &0.29593 &0.50159 &1.11220 \\
 &$MSE(\widehat{\lambda})$ &0.05064 &0.06212 &0.03012 &0.14915 &0.43178 &2.13745 \\
 &$MRE(\widehat{\lambda})$ &1.78156 &0.24306 &0.13340 &0.11837 &0.12540 &0.13902 \\
 \hline
\multirow{4}{*}{80} &$\widehat{\lambda}$ &0.14297 &0.69357 &1.00650 &2.52627 &4.04873 &8.09970 \\
 &$|Bias(\widehat{\lambda})|$ &0.13933 &0.08788 &0.08018 &0.17574 &0.30681 &0.64977 \\
 &$MSE(\widehat{\lambda})$ &0.02784 &0.01508 &0.01010 &0.05003 &0.15378 &0.69129\\
 &$MRE(\widehat{\lambda})$ &1.39332 &0.12555 &0.08018 &0.07030 &0.7670 &0.08122\\
 \hline
 \multirow{4}{*}{100} &$\widehat{\lambda}$ &0.13658 &0.69599 &1.00442 &2.52283 &4.02962 &8.06924\\
 &$|Bias(\widehat{\lambda})|$ &0.13210 &0.07578 &0.07021 &0.15601 &0.27123 &0.57281\\
 &$MSE(\widehat{\lambda})$ &0.02453 &0.01082 &0.00786 &0.03983 &0.11751 &0.53522\\
 &$MRE(\widehat{\lambda})$ &1.32095 &0.10825 &0.07021 &0.06240 &0.06781 &0.07160\\
 \hline
 \multirow{4}{*}{200} &$\widehat{\lambda}$ &0.11921 &0.69901 &1.00153 &2.51226 &4.01967 &8.04444\\
 &$|Bias(\widehat{\lambda})|$ &0.11072 &0.05340 &0.05144 &0.11073 &0.19117 &0.40153\\
 &$MSE(\widehat{\lambda})$ &0.01639 &0.00449 &0.00415 &0.01927 &0.05768 &0.25741\\
 &$MRE(\widehat{\lambda})$ &1.10715 &0.07628 &0.05144 &0.04429 &0.04779 &0.05019\\
 \hline
 \multirow{4}{*}{350} &$\widehat{\lambda}$ &0.10511 &0.69976 &1.00159 &2.50346 &4.01149 &8.02440\\
 &$|Bias(\widehat{\lambda})|$ &0.09745 &0.03960 &0.03766 &0.08534 &0.14255 &0.30229\\
 &$MSE(\widehat{\lambda})$ &0.01222 &0.00246 &0.00229 &0.01143 &0.03209 &0.14537\\
 &$MRE(\widehat{\lambda})$ &0.97452 &0.05658 &0.03766 &0.03414 &0.03564 &0.03779\\
 \hline
\end{tabular}
\end{center}
\end{table}

\section{Application}
\label{sec5}
This section encompasses the application of WRXG model to illustrate the modeling behavior on a real-life dataset. In scientific disciplines, especially in geology, the orientations of axes are of often interest. Many a time, a glaciologist is interested in measuring the orientation of the principal (or long axis) of each piece of undistributed debris during the studies for orientations of stone, pebbles or rocks deposited by a receding glacier. The data thus collected are special kind of circular data, popularly known as \textit{axial data}.\\
  
To illustrate the applicability of wrapped xgamma distribution, we analyze the Fisher-B5 dataset (Fisher,~1995) available within the $circular$ package (Agostinelli and Lund, 2017) in \textsf{R} software (version 3.5.2) (R Core Team, 2018). This dataset contains the measurements of long-axis orientation of 164 feldspar laths in basalt rock, it contains 60 orientated observations (recorded in degrees). 

\begin{figure}[!h]
\centering
\includegraphics[scale=0.7]{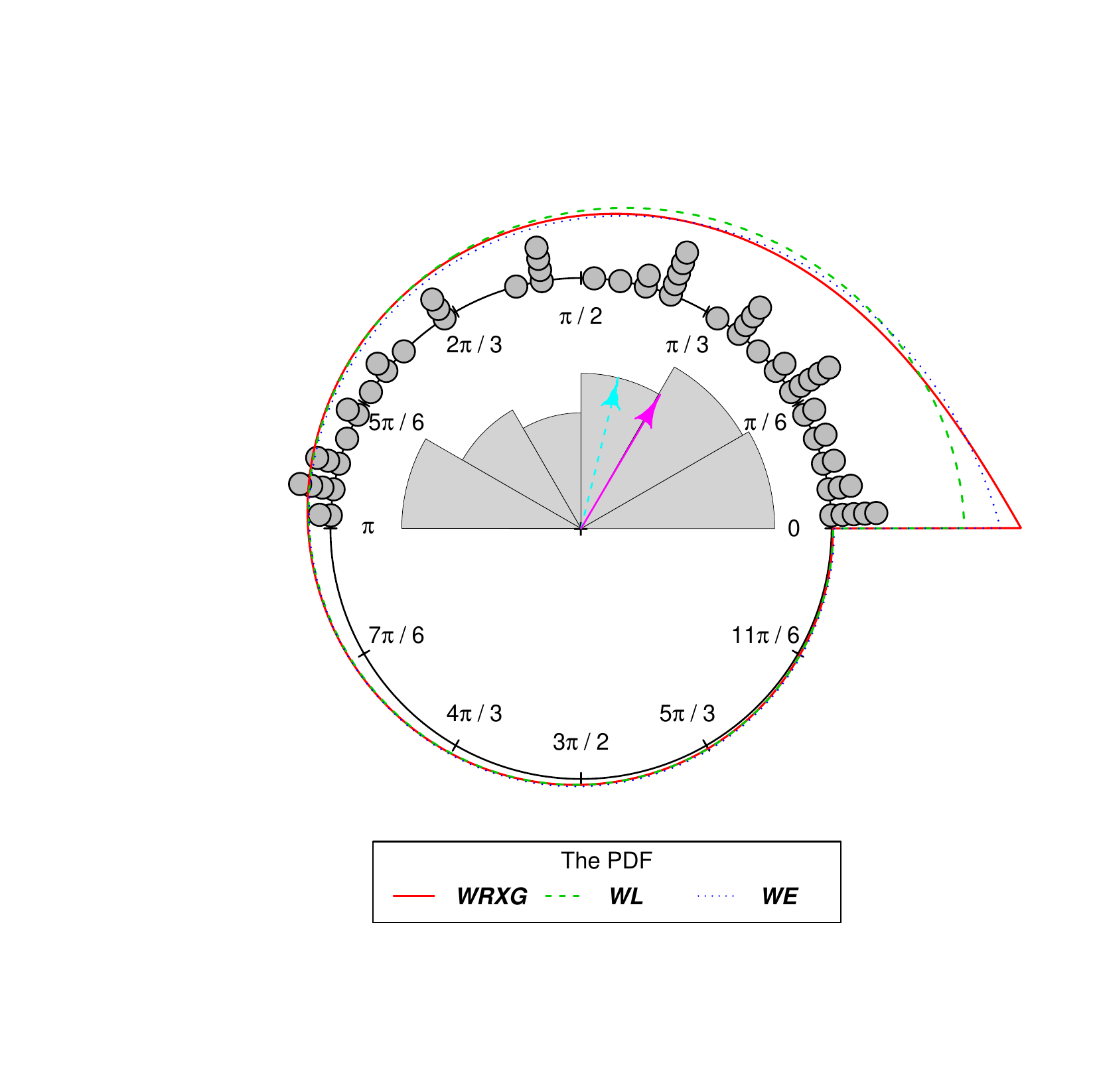}    
\caption{Plots for Fisher-B5 data. Circular data plot, fitted circular pdf and rose diagram of $WRXG(\theta;\lambda)$, $WL(\theta;\lambda)$ and $WE(\theta;\lambda)$
\label{fig:pdf_app_cir}}
\end{figure}

The ML estimation and corresponding standard error of the parameters for the Fisher-B5 data for $WRXG(\theta;\lambda)$ and two distribution, namely: wrapped Lindley distribution $WL(\theta;\lambda)$ (Joshi and Jose,~2018) and wrapped exponential distribution $WE(\theta;\lambda)$ (Jammalamadaka and Kozubowski,~2004) are obtained as given in Table~\ref{tab:tab3}. Table~\ref{tab:tab4} shows the sample mean direction, sample resultant length, mean direction, and the resultant length when the dataset is modeled with WRXG distribution. 
\begin{table}[!h]
\begin{center}
\caption{MLEs and their standard errors for the Fisher-B5 data} 
 \label{tab:tab3}
\begin{tabular}{lcc}
\hline
The Model & $\hat{\lambda}$ & SE      \\ \hline
WRXG      & 1.32075         & 0.13005 \\
WL        & 1.03085         & 0.10958 \\
WE        & 0.66400         & 0.10081 \\ \hline
\end{tabular}
\end{center}
\end{table}
In Figure~\ref{fig:pdf_app_cir}, the
dashed arrow points out the sample mean resultant vector and the solid arrow points out the mean direction
vector of the fitted WRXG distribution.\\ 
To make a comparison, Table~\ref{tab:tab5} provides the summary results of various statistics like: $-2$ log-likelihood ($-2L$), Akaike information criterion (AIC), consistent Akaike information criterion (CAIC), Bayesian information criterion (BIC), Hannan information criterion (HQIC), Durbin-Watson test statistic (W), Anderson-Darling test statistic (A) and Kolmogorov-Smirnov (K-S) and the corresponding $p$-values for the models. The smallest values of $-2L$, AIC, CAIC, BIC, HQIC, W, A, and K-S statistic with largest value of K-S $p$-value (the bold in Table~\ref{tab:tab5}), clearly show that the WRXG distribution gives the best fit to the data set comparing with the WL and WE distributions.\\ 
Figures~\ref{fig:pdf_app_cir} and~\ref{fig:denplot} provides the plots of the circular data, rose diagram, fitted pdf, estimated densities and estimated cumulative  of the fitted WRXG, WL and WE models for the given dataset.
\begin{table}[!h]
\begin{center}
\caption{ML estimates for Fisher-B5 data} 
 \label{tab:tab4}
\begin{tabular}{lcc}
\cline{2-3}
              & Mean direction & Resultant length \\ \hline
Fisher-B5 data & 1.32948       & 0.56384 ($\sim 76.17373^\circ$)        \\ \hline
WRXG Model    & 1.03693       & 0.55434 ($\sim 59.41141^\circ$)        \\ \hline
\end{tabular}
\end{center}
\end{table}

\begin{table}[!h]
\begin{center}
\caption{ML estimates for Fisher-B5 data} 
 \label{tab:tab5} \resizebox{\textwidth}{!} {\begin{tabular}{lccccccccc}
\hline
The Model  & -2$L$      & AIC      & CAIC     & BIC      & HQIC     & W        & A        & K-S(stat) & K-S($p$-value) \\ \hline
$WRXG$ & \textbf{156.0570}  & \textbf{158.0570}  & \textbf{158.1260}  & \textbf{160.1514} & \textbf{158.8763} & \textbf{0.1046} & \textbf{0.9258} & \textbf{0.1042}  & \textbf{0.5323}     \\
$WL$   & 156.8536 & 158.8536 & 158.9226 & 160.9480  & 159.6728 & 0.1199 & 1.0092 & 0.1112  & 0.4482     \\
$WE$   & 159.4301 & 161.4301 & 161.4990  & 163.5244 & 162.2493 & 0.1370 & 1.1247 & 0.1165  & 0.3900     \\ \hline
\end{tabular}}
\end{center}
\end{table}

\begin{figure}[!h]
\begin{subfigure}{.5\textwidth}
  \centering
  \includegraphics[width=0.9\linewidth]{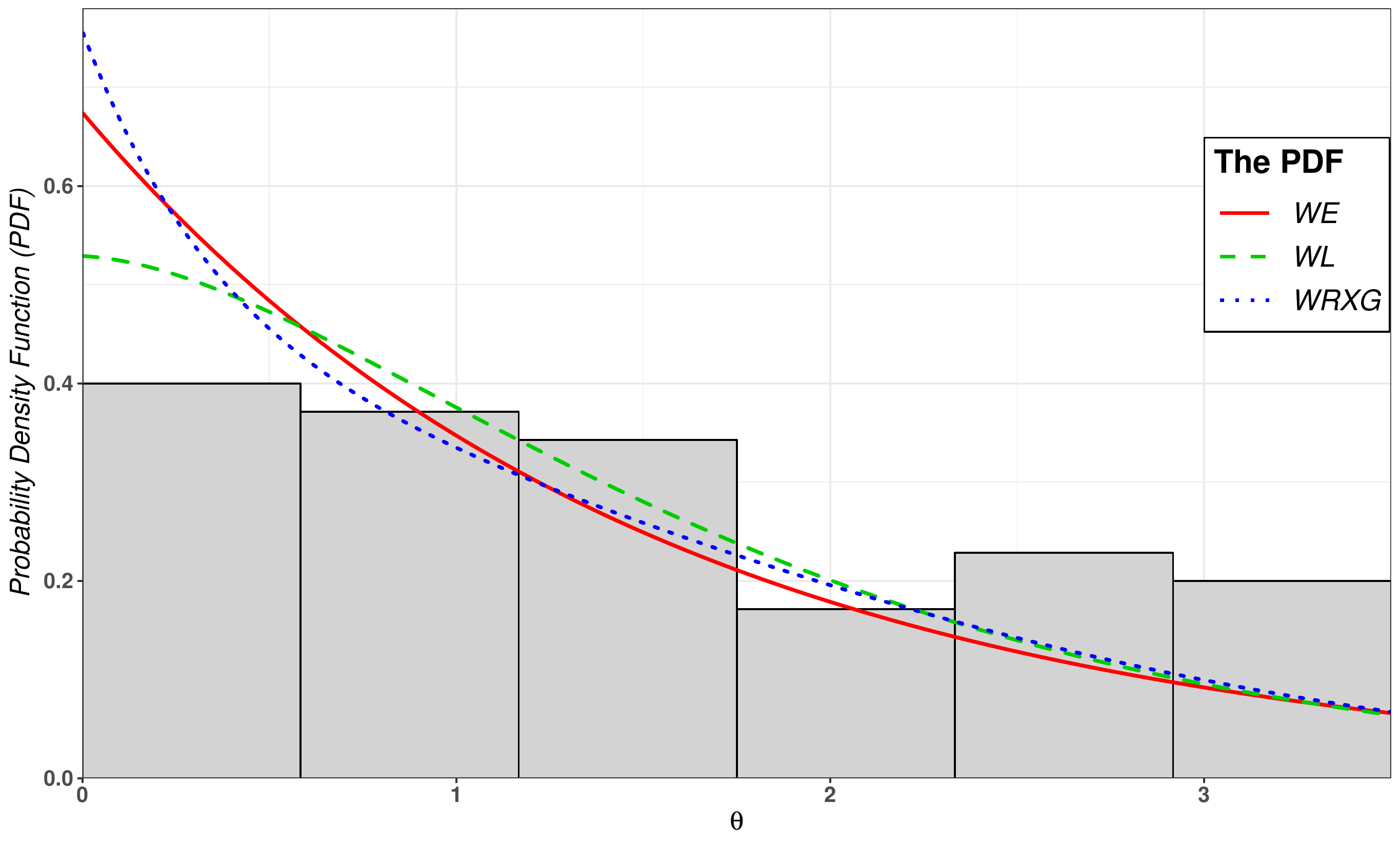}
  \caption{The linear histogram and fitted pdf.}
  \label{fig:pdf_app}
\end{subfigure}%
\begin{subfigure}{.5\textwidth}
  \centering
  \includegraphics[width=0.9\linewidth]{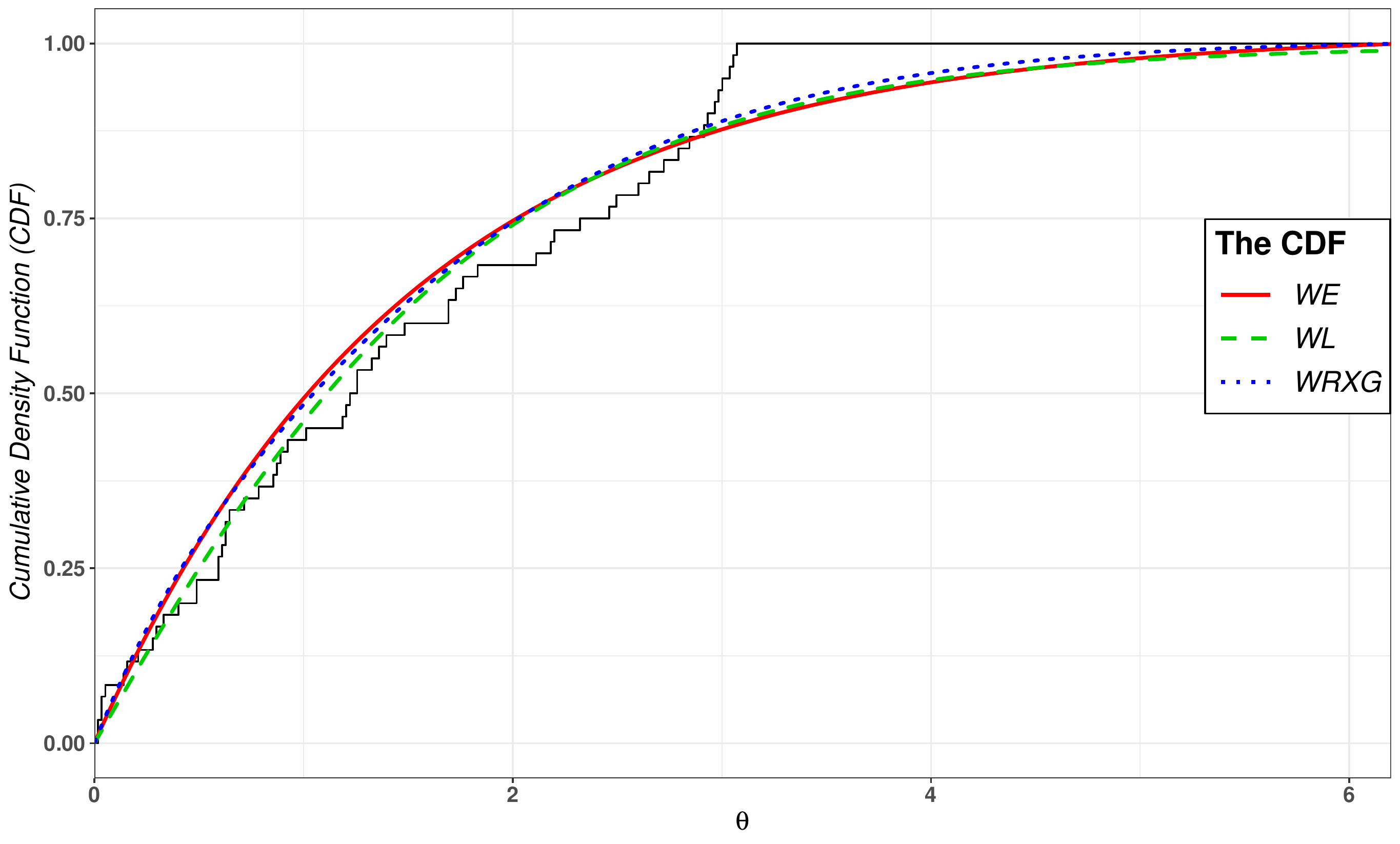}
  \caption{The empirical cdf.}
  \label{fig:cdf_app}
\end{subfigure}
\caption{The linear histogram, fitted pdf and empirical cdf of $WRXG(\theta;\lambda)$, $WL(\theta;\lambda)$ and $WE(\theta;\lambda)$, for Fisher-B5 data.}
\label{fig:denplotdata}
\end{figure}

\section{Concluding remarks}
\label{sec6}
In this article we have introduced and studied the circular version of xgamma distribution by the method of ``wrapping" and the distribution, thus obtained, is named as wrapped xgamma distribution. Different distributional properties, viz., characteristic function, trigonometric moments, mean direction, circular variance, skewness and kurtosis measures are studied for the distribution. Maximum likelihood method of estimation is been proposed for the unknown parameter and behavior of estimates have been investigated by a simulation study. The wrapped xgamma distribution found an application in modeling data on long axis orientation of feldspar laths in basalt. The proposed model came out to be quite efficient, in comparison to other circular models, for real life circular data modeling.

\end{document}